\pdfoutput=1
\documentclass[10pt, conference]{IEEEtran}
\IEEEoverridecommandlockouts
\usepackage{cite}
\usepackage{amsmath,amssymb,amsfonts}
\usepackage{algorithmic}
\usepackage{graphicx}
\usepackage{textcomp}
\usepackage{xcolor}
\usepackage{booktabs}
\usepackage{url}
\usepackage{hyperref}
\usepackage{mdframed}
\usepackage{multirow}
\usepackage{tabularray}
\usepackage{soul} 
\usepackage{needspace}

\usepackage{breakurl}

\newcounter{commentcounter}
\newboolean{showcomments}
\setboolean{showcomments}{true}
\ifthenelse{\boolean{showcomments}}{
  \newcommand{\nbc}[3]{
    \stepcounter{commentcounter}
    {\colorbox{#3}{\bfseries\sffamily\scriptsize\textcolor{white}{#1\hspace{0.1cm}\thecommentcounter}}}%
    {\textcolor{#3}{\sf\small$\blacktriangleright$\textit{#2}$\blacktriangleleft$}}}
}{
  \newcommand{\nbc}[3]{}
}

\def\BibTeX{{\rm B\kern-.05em{\sc i\kern-.025em b}\kern-.08em
    T\kern-.1667em\lower.7ex\hbox{E}\kern-.125emX}}

\begin{document}

\title{Which Prompting Technique Should I Use? An Empirical Investigation of Prompting Techniques for Software Engineering Tasks}

\author{
Enio Garcia Santana Junior\textsuperscript{1},
Gabriel Benjamin\textsuperscript{1},
Melissa Araujo\textsuperscript{1},
Harrison Santos\textsuperscript{1},
David Freitas\textsuperscript{1},\\
Eduardo Almeida\textsuperscript{1},
Paulo Anselmo da Mota Silveira Neto\textsuperscript{2},
Jiawei Li\textsuperscript{3},
Jina Chun\textsuperscript{3},
Iftekhar Ahmed\textsuperscript{3}
\\
\textsuperscript{1}Federal University of Bahia (UFBA), Brazil \\
\textsuperscript{2}Federal Rural University of Pernambuco (UFRPE), Brazil \\
\textsuperscript{3}University of California, Irvine (UCI), USA \\
}

\maketitle

\begin{abstract}
A growing variety of prompt engineering techniques has been proposed for Large Language Models (LLMs), yet systematic evaluation of each technique on individual software engineering (SE) tasks remains underexplored. In this study, we present a systematic evaluation of 14 established prompt techniques across 10 SE tasks using four LLM models. As identified in the prior literature, the selected prompting techniques span six core dimensions (Zero-Shot, Few-Shot, Thought Generation, Ensembling, Self-Criticism, and Decomposition). They are evaluated on tasks such as code generation, bug fixing, and code-oriented question answering, to name a few. Our results show which prompting techniques are most effective for SE tasks requiring complex logic and intensive reasoning versus those that rely more on contextual understanding and example-driven scenarios. We also analyze correlations between the linguistic characteristics of prompts and the factors that contribute to the effectiveness of prompting techniques in enhancing performance on SE tasks. Additionally, we report the time and token consumption for each prompting technique when applied to a specific task and model, offering guidance for practitioners in selecting the optimal prompting technique for their use cases.


    
    




    
\end{abstract}

\begin{IEEEkeywords}
Prompt Engineering; Large Language Models; Software Engineering Tasks.

\end{IEEEkeywords}

\section{Introduction}
\label{sec:intro}

Emerging Large Language Models (LLMs) have rapidly advanced beyond the capabilities of earlier smaller models, thanks to breakthroughs in deep neural architectures, access to large-scale training corpora in natural language and source code, and substantial computational power \cite{vaswani2017attention}. Simply through prompting without any model training, LLMs have already shown state-of-the-art performance in automating various Software Engineering (SE) tasks, such as code translation \cite{pan2024lost,wu2024transagents}, commit message generation \cite{li2024only,li2025consider}, and program repair \cite{ hou2024,xu2024aligning}. Despite these promising developments, it has become increasingly evident that even small changes in how a prompt is formulated can drastically alter an LLM's output \cite{sclar2024quantifying}. In automating SE tasks by prompting LLMs, the adopted prompts have the potential to impact various quality aspects ranging from code correctness and readability to the efficiency of bug-fixing suggestions\cite{The_Prompt_Report, 10.1145/3560815, wei2022cot}.

To guide LLMs toward more reliable, accurate, and context-aware responses, an array of \emph{prompt engineering} \cite{The_Prompt_Report} techniques have been proposed by researchers in recent years. However, it remains unknown which prompt engineering technique would benefit a specific SE task the most, leaving practitioners uncertain about how to compose the prompts to suit their needs. In this work, we address this gap by \emph{systematically evaluating the effectiveness of fourteen widely used prompt engineering techniques} \cite{The_Prompt_Report} within \emph{ten SE tasks} \cite{An_Empirical_Comparison}, spanning an extensive range of code understanding and generation tasks. This prompted us to ask our first research question:

\textbf{RQ1: How do different prompting techniques impact the performance of SE tasks?}

We analyzed the linguistic features of these prompts to gain a deeper understanding of the characteristics that distinguish prompting techniques that lead to performance improvements from those that do not in different SE tasks. This includes examining aspects such as Lexical Diversity, Token Count, Flesch-Kincaid Grade Level, Gunning Fog Index, and Flesch Reading Ease.
These linguistic features help identify how prompt clarity and complexity relate to LLM performance in SE tasks. Lexical Diversity reflects vocabulary richness, Token Count captures prompt length, and Flesch-Kincaid Grade Level, Gunning Fog Index, and Flesch Reading Ease assess readability and complexity. Together, they reveal whether more concise, readable, or varied prompts are associated with improved task outcomes. Identifying significant linguistic similarities among high-performing prompting strategies can reveal useful insights for systematically designing more effective prompts for SE automation. In addition to linguistic features, we investigated other potential factors contributing to the success of prompting techniques by leveraging contrastive explanation \cite{jacovi2021contrastive}. Specifically, we queried the LLMs to compare successful and less successful prompts, aiming to understand what the model perceives as key differentiators in performance. Thus, we formulated the following research questions:



\textbf{RQ2: What linguistic features of prompting techniques are associated with improved performance on SE tasks?}


\textbf{RQ3: What factors, according to LLMs, contribute to the effectiveness of prompting techniques in SE tasks?}

Since large language models (LLMs) require substantial computational resources and can incur high financial costs, particularly when accessed via commercial APIs, software practitioners must carefully balance the trade-off between performance improvements and resource consumption \cite{alizadeh2024llm}. Recent studies indicate that seemingly minor variations in prompts can result in disproportionately large differences in resource usage, including increased token counts and longer inference times. Therefore, understanding how different prompting techniques impact both performance and resource costs is essential for making informed decisions in practical SE scenarios. This motivation leads to our next research question:

\textbf{RQ4: How are resource costs associated with the performance of prompting techniques in SE tasks?}


Based on our findings, we constructed composite prompts by combining the prompting techniques that consistently yield performance improvements in SE tasks. We aimed to investigate whether integrating all positively impactful factors could lead to further performance gains. Ultimately, we aim to enable SE practitioners and researchers to leverage LLMs more effectively and accurately, advancing more robust and autonomous SE practices. This motivated our final research question:

Overall, this paper makes the following contributions:
\begin{itemize}
\item A comprehensive benchmark of \textbf{14 prompt-engineering techniques} evaluated on \textbf{10 SE tasks} and \textbf{3 LLM families}, totaling more than 2k prompts.
  
\item A \textbf{systematic investigation} that discloses the factors associated with a prompting technique's success across tasks and models.
\item A \textbf{cost-aware analysis} reporting token budgets and latency for every technique, enabling practical cost–performance trade-offs.
\item Public release of all datasets, prompts, model outputs, and analysis scripts to facilitate replication and future research.
\end{itemize}




\section{Related Work}
\label{sec:RelatedWork}

\textbf{LLMs in Software Engineering.} 
Recent LLMs have billions of parameters and are trained with large-scale datasets of natural language and source code, showing state-of-the-art understanding and generation capabilities. Consequently, researchers have increasingly adopted them to conduct various SE tasks ranging from code translation \cite{pan2024lost,wu2024transagents}, commit message generation \cite{li2024only,li2025consider}, code review \cite{guo2024exploring,sghaier2025harnessing}, and program repair \cite{hou2024,xu2024aligning} to name a few. 
In addition to research adoption, practitioners also integrate LLMs into software development activities. Recent years have witnessed numerous LLM-based AI code assistants, such as Github Copilot \cite{copilot}, JetBrains AI assistant \cite{jetbrainsAI}, and Visual Studio IntelliCode \cite{VS}. Sergeyuk et al. \cite{sergeyuk2025using} have found that developers tend to delegate certain SE tasks to those LLM-based AI code assistants, including test case and documentation generation.




\textbf{Techniques to Improve LLM Performance.}
Various techniques exist to improve the performance of LLMs when automating a task. Historically, fine-tuning with domain-specific datasets has been the go-to approach to apply the LLMs to a downstream task\cite{hanindhito2025lora}. While fine-tuning can improve a generic LLM's performance on the task, it requires heavy computational resources and curated datasets, making it an expensive and time-consuming approach. However, recent works have \cite{wei2022cot, brown2020gpt3} suggested that simply prompting the LLMs with well-designed prompt engineering techniques can be equally or more effective in certain conditions. For example, Shang et al. \cite{studyUnitTesting} compared fine-tuning and prompt-based methods for generating unit tests, showing that carefully structured prompts sometimes achieve comparable results. Similarly, Chen et al. \cite{taxonomyConstruction} demonstrated that prompting LLMs outperformed fine-tuning in taxonomy-building tasks, and the performance gap widened when the LLMs were fine-tuned with limited data. Moreover, \cite{The_Prompt_Report} offers a survey of 58 techniques, illustrating how prompt formulation can dramatically affect outcomes. This highlights the growing recognition that the way prompts are structured can significantly influence the quality of the generated output, making prompt engineering a crucial skill for developers and researchers. Despite increased interest in prompt-based methods with LLMs, best practices remain inconsistent, especially in automating software development and maintenance activities. In this study, we aim to fill this gap by providing insights into how different SE tasks would benefit from different prompting techniques so that researchers and practitioners can better utilize the power of the LLMs to automate these SE tasks.

\textbf{Comparisons of LLM Approaches.}
Prior works have investigated the effectiveness of different pre-trained LLMs in performing SE tasks. For instance,  Niu et al. \cite{An_Empirical_Comparison} evaluated 19 pre-trained models on 13 SE tasks, highlighting the correlation between LLMs of different architectures and their performance on SE tasks. Similarly, a survey on code generation \cite{surveyCodeGen} analyzed multiple LLM-based solutions and ranked their performance across standard benchmarks. Alizadeh et al. \cite{alizadeh2025energy} extended this line of work by analyzing the energy–accuracy trade-off of 18 LLM families, demonstrating that larger models and higher energy budgets do not always yield superior results and that no single model dominates across diverse SE tasks. While these comparisons have shown the importance of selecting the right model, they rarely revealed how the wording and structure of the prompt itself can influence the quality of the generated content, leading to an oversight, as even a well-trained model can underperform if not guided by well-constructed instructions formatted by an appropriate prompting technique\cite{hase2023prompting}. It remains unclear how to adapt each task-specific scenario via prompting to maximize SE task performance, an aspect we address in this work.

In addition, we note that while both fine-tuning and prompt engineering have been studied in the domain of software engineering, no comprehensive evaluation has yet examined how different prompt engineering techniques perform across SE tasks. This comparative analysis can highlight the strengths and weaknesses of various prompting techniques, offering practical guidelines for developers and researchers alike. To fill this gap, our research systematically compares recognized prompting techniques drawn from the literature on LLMs and SE across diverse scenarios such as bug fixing, test generation, and code summarization.

\textbf{Explaining the Prompt Technique.} Researchers have explored various metrics to explain model predictions and gain insights into their behavior. One such metric is \textit{contrastive explanation}, which seeks to clarify why a model predicts a specific outcome for a given input instead of an alternative outcome. In the NLP domain, prior studies have employed \textit{contrastive explanation} to better understand model decisions in tasks such as natural language inference~\cite{jacovi2021contrastive}, question answering~\cite{chen2023distinguish}, and grammatical acceptability~\cite{yin2022interpreting}. These studies demonstrate that \textit{contrastive explanations} can be more effective in revealing model reasoning than simply prompting the model for a rationale. In this work, we adopt the practices proposed in~\cite{jacovi2021contrastive} to examine, given a prompt, task instance, and model-generated output, why one prompt technique leads to better performance than another. Our work is the first to investigate \textit{why} certain prompts perform better in various SE tasks leveraging \textit{contrastive explanations}.

\section{Methodology}
In this section, we introduce the methodology of our experiments for answering our research questions.

\subsection{Software Engineering Tasks}
To make our findings more generalizable and comprehensive, we targeted a broad range of Software Engineering (SE) tasks curated by Niu et al (Table \ref{tab:se_tasks_metrics}). \cite{An_Empirical_Comparison}. These tasks have been widely explored and automated by researchers using LLMs, demonstrating various characteristics in terms of input-output modalities. They have also been grouped into code understanding and code generation tasks. Due to the limited financial budget, we could not run our experiments on all the datasets associated with these SE tasks. Thus, we conducted random sampling on each dataset with a 95\% confidence level and a 5\% margin of error.

The original study from which we derived our task set included a total of 13 tasks. For the purposes of our investigation, we refined this list to 10 tasks. Specifically, we excluded Code-to-Code Retrieval and Code Search as these tasks are primarily designed for code embedding models \cite{liu2024codexembed,voyage} rather than generative models where the design of the prompting techniques can impact the performance. Additionally, we excluded the Code Completion task due to challenges in reproducing the required dataset format using the implementation provided in the replication package \cite{An_Empirical_Comparison}. In total, we obtained 10 sampled datasets. Table \ref{tab:se_tasks_metrics} lists the sample size for each dataset for an SE task.

\begin{table*}[h!]
\centering
\caption{Summary of selected SE Tasks and Evaluation metrics used}
\label{tab:se_tasks_metrics}
\renewcommand{\arraystretch}{1.3}
\begin{tabular}{ccccc}
\toprule
\textbf{Task (Abbreviation)} & \textbf{Category} & \textbf{Dataset} & \textbf{Metric(s)} & \textbf{Sample size} \\
\midrule
Defect Detection (DD) & Code Understanding & Devign~\cite{Learning_and_evaluating} & Acc & 391 \\
Clone Detection (CD) & Code Understanding & BigCloneBench~\cite{bigclonebench} & F1 & 390 \\
Exception Type Prediction (ET) & Code Understanding & Kanade et al.~\cite{Learning_and_evaluating} & Acc & 380 \\
Code Question Answering (QA) & Code Understanding & CoSQA~\cite{cosqa}, CodeSearchNet~\cite{codesearchnet} & Acc & 382 \\
Code Translation (CT) & Code Generation & CodeTrans~\cite{codesearchnet} & CodeBLEU~\cite{CodeBLEU} & 378 \\
Bug Fixing (BF) & Code Generation & BFP~\cite{bfp} & CodeBLEU~\cite{CodeBLEU} & 390 \\
Mutant Generation (MG) & Code Generation & GM~\cite{gm} & BLEU~\cite{Papineni02bleu} & 390 \\
Assert Generation (AG) & Code Generation & ATLAS~\cite{atlas} & BLEU~\cite{Papineni02bleu} & 390 \\
Code Summarization (SM) & Code Generation & DeepCom~\cite{codesearchnet} & BLEU~\cite{Papineni02bleu} & 390 \\
Code Generation (CG) & Code Generation & CONCODE~\cite{concode} & CodeBLEU~\cite{CodeBLEU} & 391 \\
\bottomrule
\end{tabular}
\end{table*}


\subsection{Prompting Techniques}\label{sec:Defining_Prompting_Techniques}
A prompting technique is a blueprint that describes how to structure a prompt that serves as the input query to the LLMs. It may incorporate conditional logic, parallelism, or other architectural considerations to obtain responses from the LLMs that better align with the user intent and task objectives \cite{The_Prompt_Report}.


Schulhoff et al. \cite{The_Prompt_Report} curated a catalog of 46 prompting techniques in literature. To ensure that prompting techniques can be readily applied to the selected SE tasks, we devised a set of exclusion criteria to filter out those that were not relevant or applicable. This filtering process was conducted independently by two researchers, master's students with more than two years of experience in prompt engineering. In cases where disagreement arose, a third researcher with extensive experience in LLMs and software development was consulted to resolve the divergence and reach a consensus. The whole process was conducted through a negotiated agreement \cite{forman2007qualitative}. These filtering criteria are: (1) Only one prompting technique was retained if multiple prompting techniques function similarly (i.e,. K-Nearest Neighbor\cite{liu-etal-2022-makes} and Vote-K \cite{su2022selectiveannotationmakeslanguage} both are few-shot techniques that differ only on the method used to choose the examples); (2) the prompting technique cannot be an ensemble of multiple other techniques (i.e,. DENSE \cite{khalifa2023exploring}); (3) Since our goal was to examine the effectiveness of the prompt content and structure, the prompting technique cannot rely on external task-specific tools (i.e., ReAct \cite{yao2023react}).

After filtering, 14 out of the 46 prompting techniques were adopted in our analysis. The list of all 46 prompting techniques and the rationale for whether they were included or not is available in our replication package \footnote{\url{https://github.com/prompt-study/prompt-tasks-study/tree/main}}. We describe each of the 14 techniques as follows:


\begin{enumerate}
    \item \textbf{Exemplar Selection KNN\cite{liu-etal-2022-makes} (ES-KNN):} Selects exemplars using a k-nearest neighbor approach to enrich the prompt context. In detail, the data samples from the full original dataset (excluding the current data sample to be queried) are transformed into vector embeddings using a code-optimized embedding model\cite{jina}. We selected the most widely adopted and top-performing embedding model available at the time of our experiment. Depending on the SE task, since the data samples may consist of source code, natural language, or both, we selected a multilingual embedding model that was trained with both natural language and multiple programming languages. We then computed cosine similarity between the current sample and encoded data samples to retrieve the top-k nearest neighbor samples, which were selected as in-context exemplars to be added to the prompt. We selected the most freqency used and best performing 
    
     
    \item \textbf{Few Shot Contrastive CoT\cite{chia2023contrastivechainofthoughtprompting} (CCoT):} Uses both correct and incorrect chain-of-thought examples to refine reasoning steps.
    \item \textbf{Tree Of Thought\cite{yao2023treethoughtsdeliberateproblem} (TroT):} Structures multiple branching reasoning paths to explore diverse solution strategies for complex design problems.
    \item \textbf{Self Ask\cite{press2023measuringnarrowingcompositionalitygap} (SA):} Makes the model generate its own follow-up questions before answering, which helps it break down and solve complex problems step by step. 
    \item \textbf{Universal Self Consistency\cite{chen2023universalselfconsistencylargelanguage} (USC):} Combines multiple answers from the model and uses a meta-prompt to pick the most consistent one, improving output reliability.
    
    \item \textbf{Self Refine\cite{madaan2023selfrefineiterativerefinementselffeedback} (SR):} Iteratively improves initial responses by self-evaluating and updating code explanations or solutions.
    \item \textbf{Self-Generated In-Context Learning\cite{kim2022selfgeneratedincontextlearningleveraging} (SG-ICL):} Automatically generates in-context exemplars to simulate few-shot learning, streamlining prompt formulation for coding tasks.
    \item \textbf{Thread Of Thought\cite{zhou2023threadthoughtunravelingchaotic} (ToT):} This technique guides the LLMs to work through a problem step by step, focusing on breaking down a large or complex task into smaller, manageable parts. For example, instead of solving everything at once, the model is told to pause, summarize, and analyze each step before moving to the next step, making the reasoning process more organized and clear.
    \item \textbf{Step Back Prompting\cite{zheng2024stepbackevokingreasoning} (SBP):} With step back prompting, the model is first asked to examine the problem as a whole and think about the key ideas or main facts, before drafting the solution. This helps the LLMs to plan ahead and avoid jumping into details.
    
    \item \textbf{Emotional Prompting\cite{li2023largelanguagemodelsunderstand} (EP):} Incorporates affective language to shape engaging and empathetic responses, useful in writing user-friendly documentation or error messages.
    \item \textbf{Style Prompting\cite{lu2023boundingcapabilitieslargelanguage} (SP):} Directs the model to adopt a specific tone or format, ensuring that generated code comments and documentation align with a desired style.
    \item \textbf{Rephrase and Respond\cite{deng2024rephraserespondletlarge} (RR):} This technique asks the LLMs to first restate the question in its own words and add any extra details if needed, before giving an answer. By making the LLM explain the question to itself, the model is assumed to be more likely to fully understand what is being asked and give a more accurate and detailed response. 
    
    \item \textbf{Role Prompting\cite{wang2024rolellmbenchmarkingelicitingenhancing} (RP):} Assigns a specific persona---such as a code reviewer or developer---to tailor responses to the nuances of specific SE tasks.
    \item \textbf{Analogical Prompting\cite{yasunaga2024largelanguagemodelsanalogical} (AP):} This technique prompts the LLMs to use analogies to make code explanations or design ideas easier to understand. This helps turn complex or abstract data structures or algorithms into relatable, real-world examples.
\end{enumerate}

\subsection{Prompt Validation}\label{sec:prompt_validation}

\begin{figure}[htbp]
    \centering
    \includegraphics[width=0.5\textwidth]{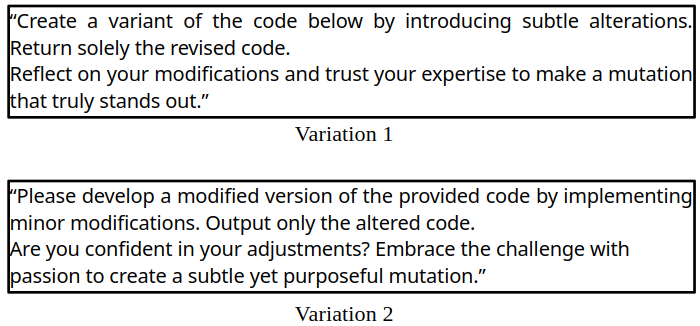}
    \caption{Example of two variations of the emotion prompting template for the task Mutant Generation.}
    \label{fig:variations}
\end{figure}

We conducted a structured prompt validation process~\cite{schreiter2024prompt} to reduce the potential impact of prompt wording on the task performance. For each prompting technique, we constructed ten prompt variation templates with different synonyms and phrasings, using OpenAI's ChatGPT\cite{chatgpt} as a tool to ensure no loss in text quality\cite{investigating_the_acc, is_chatgpt_a_highly} (e.g., eliminating grammatical errors, maintaining sentence coherence, and preserving the original intent of the technique), as well as to maximize variability across the generated prompts. Figure ~\ref{fig:variations} contains two examples of prompt template variations. Then six researchers, who were also the authors of this study, participated in the manual process of reviewing these templates to ensure that the semantic meaning of the prompts remained consistent. To detail, a pair of researchers was assigned 4 to 5 prompting techniques. Each researcher independently reviewed the variation templates for their assigned technique to determine whether the prompts conformed to the descriptions provided by scientific literature. A variation template was accepted only if both researchers in the pair agreed. In cases of disagreement, the template was discarded, and new variation was generated and reviewed until ten acceptable template variations were finalized.

The average Cohen's kappa for inter-rater reliability across all prompting techniques was \textit{kappa=0.45}.

\subsection{Language Model Selection}
To ensure the generalizability and comprehensiveness of our findings, we employed a diverse set of LLMs. These models have been developed by different organizations, focusing on various architectures and training objectives. Since the selected SE tasks involve not only natural language but also source code, we chose the LLMs that show state-of-the-art performance on widely adopted code generation benchmarks, such as EvalPlus \cite{10.5555/3666122.3667065}. Consequently, we selected \textit{DeepSeek-V3} \cite{deepseekai2025deepseekv3technicalreport}, \textit{Qwen2.5‑Coder‑32B‑Instruct} \cite{hui2024qwen25codertechnicalreport}, \textit{Llama-3.3‑70B‑Instruct} \cite{meta2024llama33}, and \textit{OpenAI o3-mini}\cite{openai2025o3mini}. 



\subsection{Result Collection}
\label{sec:result_collection}

The results were obtained by applying each prompting technique to all ten selected SE tasks (Table \ref{tab:se_tasks_metrics}). For each prompting technique being applied to an SE task, we divided the dataset evenly and ensured all prompt template variations were applied as close to the same number of data instances as possible (between 38 and 40 instances per variation). These prompt template variations (Section~\ref{sec:prompt_validation}) were then sent to the LLMs for solving the SE task. Specifically we used together.ai API \cite{together} for the open source language models: \textit{DeepSeek-V3}, \textit{Qwen2.5‑Coder‑32B‑Instruct} and \textit{Llama-3.3‑70B‑Instruct}, and for the \textit{o3-mini} we used the OpenAi API. The raw model responses were subsequently processed using task-specific extraction scripts. Our parser aimed to extract only the relevant answer field from the model’s output according to the expected output format for each SE task. To ensure the parser did not artificially inflate performance, we only extracted answers delimited by special markers that were specified in the prompt instructions. If the model correctly used the specified delimiters, our parser would return only the content within those delimiters. If the model response failed to include the required delimiters, we included the entire response in our analysis. We included the entire model response as the answer for evaluation. This process ensured that the extraction step reflected the actual performance of the LLMs in adhering to prompt specifications and instructions. After this cleaning step, the extracted responses were used to compute each SE task's evaluation metrics (Table \ref{tab:se_tasks_metrics}).



In addition, we constructed a baseline to demonstrate the effectiveness of the prompting techniques under study. Specifically, the baseline only prompts the LLMs with a simplistic instruction to complete the SE task along with the data sample. For each of the ten SE tasks, we averaged performance metrics across the four LLMs to have an overview of the effectiveness of the prompting techniques. Then, we ranked the prompting techniques to identify the most effective one for the SE tasks. The ranking was determined by aggregating the performance metric for each technique across four LLMs. The metrics for the SE tasks are included in Table \ref{tab:se_tasks_metrics}. We describe them in the following:

\subsection{Linguistic Metrics}
\label{sec:linguistic_metrics}
To reveal characteristics that have the potential to impact the performance of the SE tasks when prompting LLMs, we selected the following linguistic metrics that are relevant to prompt engineering: 



\begin{itemize}
    \item \textbf{Lexical Diversity (MATTR)} assesses the variety of vocabulary used in the prompt by calculating the Moving-Average Type-Token Ratio (MATTR) \cite{covington2010mattr}.
    \item \textbf{Flesch-Kincaid Grade Level} evaluates the readability of a prompt by estimating the U.S. school grade level necessary for its comprehension \cite{kincaid1975readability}.
    \item \textbf{Gunning Fog Index} calculates the years of formal education a reader needs to understand the prompt \cite{gunning1952technique}. 
    \item \textbf{Token Count} calculates the number of tokens in the prompt.
    \item \textbf{Flesch Reading Ease} rates prompt readability on a scale from 0 (very difficult) to 100 (very easy), with higher scores indicating easier-to-read prompts. This metric is widely used in software documentation and requirements engineering studies \cite{flesch1948new,LEHNER1993133}. 
    

\end{itemize}

We calculated these metrics for all the prompts in this study and averaged the metric values of all the prompts using a prompting technique for each of the selected SE task. The goal was to collect the linguistic characteristics of different prompting techniques for the SE tasks. To investigate their association with the prompting techniques' effectiveness in the SE tasks, we conducted Spearman correlation ~\cite{spearman1904proof} between the averaged values of these linguistic metrics and the performance scores of the SE tasks (Table \ref{tab:se_tasks_metrics}). 




\subsection{Contrastive Explanation}\label{sec:contrastive_explanation}

To further investigate why a specific prompting technique shows better performance in an SE task than other techniques, we used contrastive explanation \cite{jacovi2021contrastive} that attempts to uncover the factors associated with the model's decisions. Contrastive explanations provide interpretable results similar to what a human would explain a decision, which inherently answers the question \textit{``why P, rather than Q?''} \cite{hilton1988logic}. Since the decisions the LLMs make are often complex, the complete explanations can be complicated and uninterpretable \cite{jacovi2020towards}. Contrastive explanations omit the causal attributes or factors common to both P and Q, making the explanations easy to understand \cite{jacovi2021contrastive}. 


To utilize contrastive explanations to explain the superiority of a prompting technique in solving the SE tasks, we prompted the LLMs using in-context learning \cite{gao2023makes}. To be more specific, for each SE task, we included the definitions of the prompting technique that performed the best and the worst (Section \ref{sec:result_collection}) and prompted the LLM to generate contrastive explanations. In total, we prompted the LLM (i.e., the same LLM that had generated the responses for the SE task) to obtain comprehensive explanations regarding the potential factors that were responsible for a prompting technique's superior performance compared to worst prompting technique. The process was repeated for all four LLM models used in our experiment.

To ensure the quality and interpretability of the explanations, we also attached four demonstration examples (following the best practice of in-context learning in software engineering research \cite{gao2023makes}), each consisting of the queries constructed by the two prompting techniques and their responses. To make the responses of the two prompting techniques more distinguishable for the LLM to generate explanations, we only included examples whose responses by the superior prompting technique were significantly better than those of the worst prompting technique (For understanding tasks, the response of the superior prompting technique is a correct prediction while that of the compared technique is not. For generation tasks, we ensure the metric scores of the two techniques have the most difference). Due to the space limit, we include the complete prompt for all the SE tasks in our replication package. To ensure the quality of the prompt to extract explanations, we followed the best practices \cite{goriparthi6202advancing} and examined the responses after executing the composed prompts ten times. 

We collected all the contrastive explanation responses provided by the four LLM models. Then, for each SE task, the responses across the models were combined for manual evaluation. To detail, we used a card sorting method \cite{zimmermann2016card} to conduct a qualitative analysis of these responses. Two researchers participated in this analysis. Each of them evaluated the explanations and categorized them into codes. They then met to discuss these codes, which were further grouped into high-level categories. Finally, the researchers refined the categories and organized them into meaningful themes that can sufficiently explain why a certain technique outperformed the others.  

\section{Results}
\label{sec:Results}


\subsection{(RQ1) How do different prompting techniques impact the performance of SE tasks?}\label{sec:RQ1} 

Table~\ref{tab:best_prompting_techniques} shows the best-performing prompting technique for each SE task using different LLMs. To have an overview of the effectiveness of the prompting techniques for each task, we also provided the prompting techniques that showed the best performance, averaged across all the LLMs in the \textit{Aggregate} column. We note that while no single prompting technique emerged as universally optimal across all tasks, there are certain prompting techniques that tend to show superior effectiveness than others. For example, for tasks such as \textit{Clone Detection}, \textit{Code Translation}, and \textit{Assert Generation}, \textit{ES-KNN} consistently outperforms other prompting techniques across all four selected LLMs. This suggests that semantically similar in-context demonstration examples in the prompt may be able to hint the LLMs how to process the current data sample to achieve the goals of the SE tasks properly. In addition, \textit{USC} demonstrates the best performance in \textit{Code QA} and \textit{Code Generation} for most of the LLMs, which indicates that providing the model with structured examples or encouraging exploration of multiple solution pathways can mitigate errors in the generated output. It is worth noting that \textit{ToT} presents outstanding performance for the task of \textit{Defect Detection}. It prompts the LLMs to engage in a process where the code is broken down into components, and each component is examined step by step to identify the potential defects. Such a process may push the LLMs to focus on specific components one at a time, increasing the chance of identifying defects. 

As for the four selected LLMs, \textit{DeepSeek-V3}, \textit{Qwen2.5-Coder-32B-Instruct}, and \textit{Llama-3.3-70B} show nearly consistent results regarding the prompting technique that is the most effective across all SE tasks (i.e., \textit{ES-KNN} is the most effective one across all these three models in \textit{Clone Detection}, \textit{Exception Type Prediction}, \textit{Code Translation}, and \textit{Assert Generation}). However, the outperforming prompting techniques that are applied to \textit{OpenAI o3-mini} are different for most SE tasks, with \textit{RP} being the majority winner.

We also include the worst-performing prompting techniques for the SE tasks in Table \ref{tab:worst_prompting_techniques}. Table \ref{aggregate_performance_data} lists the prompting techniques along with the average values of the evaluation metrics across the four LLMs for the ten SE tasks in this study. We also included the aggregate performance of the two task groups: code understanding and generation tasks, which are displayed in z-score\cite{Iverson2011} format to make comparing different metrics possible. Here, a z-score indicates how many standard deviations a particular value is from the mean, allowing for a standardized comparison of performance across different metrics and scales.

It shows that the worst-performing prompting techniques even underperform the baseline (Section \ref{sec:result_collection}) where simplistic instruction is used. This finding points out that complex prompting techniques would not necessarily improve performance when applied to achieve certain goals. Our results reveal compatibilities of various prompting techniques to different SE tasks, providing practical guidelines to researchers and software developers. These guidelines can be essential to ensure software quality since LLMs are increasingly being used to automate software development and maintenance activities \cite{sergeyuk2025using}.

\begin{mdframed}[backgroundcolor=white, linewidth=1pt, linecolor=black]
    \textbf{Observation 1}:
No single prompt technique consistently outperforms others across all SE tasks. Certain prompting techniques can negatively impact their performance.
\end{mdframed}


\begin{table}
\centering
\footnotesize
\caption{Best Prompting Techniques for each model and Aggregate}
\label{tab:best_prompting_techniques}
\renewcommand{\arraystretch}{1.3}
\resizebox{\linewidth}{!}{
\begin{tabular}{lcccccc}
\toprule
\textbf{Task} & \textbf{Aggregate} & \textbf{Qwen} & \textbf{DeepSeek} & \textbf{Llama} & \textbf{o3-mini} \\
\midrule
\textbf{Understanding Tasks} & ES-KNN & ES-KNN & ES-KNN & ES-KNN & RP \\
Defect detection & ToT & ToT & ToT & ToT & RP \\
Clone detection & ES-KNN & ES-KNN & ES-KNN & ES-KNN & ES-KNN \\
Exception type & ES-KNN & ES-KNN & ES-KNN & ES-KNN & RR \\
Code QA & USC & SG-ICL & USC & USC & RP \\
\midrule
\textbf{Generation Tasks} & ES-KNN & ES-KNN & ES-KNN & ES-KNN & ES-KNN \\
Code translation & ES-KNN & ES-KNN & ES-KNN & ES-KNN & ES-KNN \\
Bug fixing & Control & SG-ICL & SG-ICL & SG-ICL & Control \\
Mutant generation & ES-KNN & RP & ES-KNN & ES-KNN & RP \\
Assert generation & ES-KNN & ES-KNN & ES-KNN & ES-KNN & ES-KNN \\
Code summarization & Control & SG-ICL & ES-KNN & Control & SG-ICL \\
Code generation & USC & USC & SG-ICL & USC & USC \\
\bottomrule
\end{tabular}
}
\end{table}

\begin{table}
\centering
\caption{Worst Prompting Techniques for each model and Aggregate}
\label{tab:worst_prompting_techniques}
\renewcommand{\arraystretch}{1.3}
\resizebox{\linewidth}{!}{
\begin{tabular}{lcccccc}
\toprule
\textbf{Task} & \textbf{Aggregate} & \textbf{Qwen} & \textbf{DeepSeek} & \textbf{Llama} & \textbf{o3-mini} \\
\midrule
\textbf{Understanding Tasks} & SR & SR & CCoT & SR & SBP \\
Defect detection & EP & RR & EP & EP & SBP \\
Clone detection & SR & SG-ICL & SG-ICL & USC & SR \\
Exception type & RR & SR & ToT & SR & SR \\
Code QA & TroT & TroT & TroT & Control & Control \\
\midrule
\textbf{Generation Tasks} & SG-ICL & SG-ICL & ToT & SG-ICL & SR \\
Code translation & SG-ICL & SG-ICL & USC & TroT & USC \\
Bug fixing & RR & ToT & RR & SR & SA \\
Mutant generation & ToT & ToT & SG-ICL & USC & SR \\
Assert generation & SR & TroT & SG-ICL & USC & SR \\
Code summarization & RR & RR & RR & SA & RR \\
Code generation & ES-KNN & TroT & TroT & TroT & ES-KNN \\
\bottomrule
\end{tabular}
}
\end{table}

\begin{table*}
\centering
\footnotesize
\caption{Comparison of Prompting Techniques with Associated Reason Categories}
\label{aggregate_performance_data}
\renewcommand{\arraystretch}{1.3}
\begin{tabular}{p{3cm}lcccccp{5cm}}
\toprule
  &
  & \multicolumn{2}{c}{\textbf{Best}} 
  & \textbf{Control Prompt} 
  & \multicolumn{2}{c}{\textbf{Worst}} 
  & \\
\cmidrule(lr){3-4}\cmidrule(lr){5-5}\cmidrule(lr){6-7}
\textbf{Task} & Std & Technique & z-score & z-score & Technique & z-score & Best vs Worst Contrastive Explanation\\
\midrule
\textbf{Understanding Tasks} &  & ES-KNN & 1.12 & 0.29 & ToT & -0.66 \\
Defect detection & 3.60 & ToT & 73.66 & 66.22 & SR & 59.91 & Structured Guidance, Robustness\\
Clone detection & 1.17 & ES-KNN & 68.60 & 66.03 & SG-ICL & 63.45 & Structured Guidance, Efficiency\\
Exception type & 7.16 & ES-KNN & 82.50 & 78.16 & ToT & 52.04 & Structured Guidance, In-Context Example\\
Code QA & 1.39 & USC & 55.67 & 50.99 & TroT & 51.44 & Ambiguity Reduction, Structured Guidance\\
\midrule
\textbf{Generation Tasks} &  & ES-KNN & 1.83 & 0.80 & TroT & -2.11 \\
Code translation & 6.30 & ES-KNN & 42.08 & 30.19 & TroT & 13.39  & Structured Guidance, In-Context Example\\
Bug fixing & 4.69 & SG-ICL & 36.02 & 36.26 & TroT & 16.45 & In-Context Example, Robustness\\
Mutant generation & 19.59 & ES-KNN & 69.93 & 67.98 & TroT & 16.44 & Structured Guidance, In-Context Example\\
Assert generation & 15.23 & ES-KNN & 65.44 & 25.24 & TroT & 0.92 & Structured Guidance, In-Context Example\\
Code summarization & 1.51 & SG-ICL & 4.15 & 4.16 & ToT & 0.45 & Structured Guidance, In-Context Example\\
Code generation & 3.04 & USC & 24.44 & 23.18 & TroT & 13.45 & Ambiguity Reduction, Efficiency\\
\bottomrule
\end{tabular}
\end{table*}

\subsection{(RQ2) What linguistic features of prompting techniques are associated with improved performance on SE tasks?}
\label{sec:rq2}

To examine how linguistic characteristics of prompts relate to task performance, we analyzed selected linguistic metrics (Section \ref{sec:linguistic_metrics}) across various task groups. Note that the tasks are grouped into code understanding tasks and code generation tasks following \cite{An_Empirical_Comparison} (Table \ref{tab:se_tasks_metrics}). Our goal was to identify potential associations between these features and SE task performance. The results are summarized below:



Lexical diversity, measured by the Moving-Average Type-Token Ratio (MATTR), shows a strong positive correlation with performance across all tasks (aggregate: $r=0.4440$, $p<0.001$). This correlation is particularly pronounced in code generation tasks ($r=0.5229$, $p<0.001$) and also significant in code understanding tasks ($r=0.3468$, $p=0.0088$). These results suggest that prompts with a rich vocabulary and a diverse use of words can enhance model performance.

Token count exhibits a negative correlation with performance in both code understanding ($r=-0.2567$, $p=0.0022$) and code generation tasks ($r=-0.3200$, $p=0.0030$). This indicates that longer prompts may not necessarily lead to better outcomes.

Readability metrics show varying correlations depending on the SE tasks. In code understanding tasks, Flesch-Kincaid Grade Level scores correlate negatively with task performance ($r=-0.2974$, $p=0.0260$), suggesting that simpler prompts are more effective. However, in code generation tasks, there's a positive correlation ($r=0.2975$, $p=0.0060$), implying that more complex prompts may benefit these tasks.

Similarly, the Gunning Fog Index shows a negative correlation in code understanding tasks ($r=-0.3795$, $p=0.0039$) and a positive correlation in code generation tasks ($r=0.2938$, $p=0.0067$).

However, The Flesch Reading Ease score shows different correlations. It correlates positively with task performance in code understanding tasks ($r=0.2797$, $p=0.0369$) and negatively with code generation tasks ($r=-0.3366$, $p=0.0017$). This would indicate that readability impacts performance differently depending on the task characteristics, but different metrics demonstrate different correlation patterns. We argue that future research should further investigate how the readability of the prompts impacts the performance of the LLMs in the SE tasks. Therefore, our results underscore the importance of tailoring prompt characteristics to the specific nature of the task. Enhancing lexical diversity appears universally beneficial, while optimizing readability requires further investigation.

\begin{mdframed}[backgroundcolor=white, linewidth=1pt, linecolor=black]
\textbf{Observation 2}: Linguistic features of prompts exhibit significant and various correlations with SE task performance.
\end{mdframed}

\subsection{(RQ3) What factors, according to LLMs, contribute to the effectiveness of prompting techniques in SE tasks\label{sec:rq3}}

To understand why a specific prompting technique outperforms others in SE tasks, we used contrastive explanations to examine the additional factors contributing to the effectiveness of each technique. Table~\ref{aggregate_performance_data} presents the categorized factors of the best prompting technique in contrast to the worst techniques, which was manually evaluated as outlined in~\ref{sec:contrastive_explanation}. For example, in the code translation task, the best prompting technique (\textit{ES-KNN}) included structured guidance and in-context examples, in contrast to the worst technique (Three-of-Thought). Categories of factors in Table~\ref{aggregate_performance_data} represent additional factors incorporated into the best prompting technique that contributed to superior task performance. The findings from the contrastive explanation analysis suggest that \textit{Structured Guidance} and \textit{In-Context Examples} are the most prevalent factors among the best-performing techniques. Furthermore, these common factors consistently appear across the majority of the SE tasks included in our experiment. This indicates that, for most SE tasks, effective prompts that enhance task performance tend to provide the model with structured guidance and relevant in-context examples. Below is the list of all categories for the factors identified for the best prompting techniques based on the manual analysis explained in Section~\ref{sec:contrastive_explanation}

\begin{itemize}
    \item \textbf{Ambiguity Reduction (15.38\%):} Instructions to clarify task requirements.
    \item \textbf{Correctness and Precision (2.56\%):} Instructions to ensure relevant and accurate task output.
    \item \textbf{Efficiency (17.95\%):} Instructions to deduce an optimal solution.
    \item \textbf{In-Context Example (21.80\%):} Contains examples relevant to the task.
    \item \textbf{Reasoning (2.56\%):} Instructions to include detailed justifications to support the response.
    \item \textbf{Robustness/Comprehensiveness (7.70\%):} Instructions to ensure implementations satisfy all task-specific requirements.
    \item \textbf{Structured Guidance (32.05\%):} Includes instructions to follow the conventions and patterns aligned with task output.
\end{itemize}

\begin{mdframed}[backgroundcolor=white, linewidth=1pt, linecolor=black]
    \textbf{Observation 3}:
    Prompting techniques show greater effectiveness when they include structured guidance aligned with task objectives and relevant examples.
\end{mdframed}

\subsection{{(RQ4) How are resource costs associated with the performance of prompting techniques in SE tasks?}\label{sec:rq4}}

To investigate how the resource costs impact the performance of the prompting techniques, we examine how resource-efficient each prompting technique is in conducting the SE tasks. To do so, we considered two resource measurements, namely, the number of tokens in the prompt and the response time. We normalize the efficiency of each prompting technique by dividing the values of the performance evaluation metrics by the number of used tokens and the response time. Finally, the prompting techniques are ranked based on the resource-efficiency. We present the most and least resource-efficient prompting techniques in Table \ref{tab:best_cost_benefit_prompting_techniques} and \ref{tab:worst_cost_benefit_prompting_techniques}, along with the original best and worst prompting techniques in Section \ref{sec:RQ1}.


Our findings reveal that, while \textit{ES-KNN} remains at the top for many tasks in Section \ref{sec:RQ1}, it is not the most token-efficient one. The additional context and examples provided by \textit{ES-KNN} increase token consumption, making it less suitable in settings where minimizing token usage is critical. In contrast, RP consistently ranks at or near the top for token efficiency across most tasks.  

\begin{table}
\centering
\caption{Mean number of tokens saved per prompt}
\label{tab:tokens_saved}
\renewcommand{\arraystretch}{1.3}
\resizebox{\linewidth}{!}{
\begin{tabular}{lcccccc}
\toprule
\textbf{Task} & \textbf{Aggregate} & \textbf{Qwen} & \textbf{DeepSeek} & \textbf{Llama} & \textbf{o3-mini} \\
\midrule
Defect detection & 2361.47 & 2195.55 & 3079.72 & 4867.86 & 2559.22 \\
Clone detection & 3588.50 & 3518.13 & 3777.58 & 3498.83 & 3634.21 \\
Exception type & 1985.42 & 1957.17 & 2070.51 & 1913.72 & 2583.11 \\
Code QA & 2143.98 & 2812.42 & 3984.25 & 19485.09 & 1696.93 \\
Code translation & 3021.62 & 3094.46 & 3265.77 & 3471.79 & 551.83 \\
Bug fixing & 8306.71 & 7698.30 & 8744.93 & 7621.01 & 360.08 \\
Mutant generation & 2945.01 & 92.55 & 2679.70 & 4458.38 & 362.61 \\
Assert generation & 250.75 & 427.77 & 3886.71 & 5269.65 & 643.66 \\
Code summarization & 3129.23 & 3408.88 & 4586.85 & 2648.01 & 1004.58 \\
Code generation & 10733.42 & 9278.82 & 8502.98 & 8837.50 & 791.71 \\
\bottomrule
\end{tabular}
}
\end{table}
\begin{table}
\centering
\caption{Mean value of time (seconds) saved per prompt}
\label{tab:time_saved}
\renewcommand{\arraystretch}{1.3}
\resizebox{\linewidth}{!}{
\begin{tabular}{lcccccc}
\toprule
\textbf{Task} & \textbf{Aggregate} & \textbf{Qwen} & \textbf{DeepSeek} & \textbf{Llama} & \textbf{o3-mini} \\ 
\midrule
Defect detection & 107.60 & 220.09 & 103.48 & 56.87 & 66.87 \\
Clone detection & 80.31 & 191.34 & 83.02 & 30.81 & 32.45 \\
Exception type & 66.68 & 186.78 & 22.89 & 19.91 & 54.61 \\
Code QA & 85.73 & 207.76 & 68.15 & 39.70 & 43.34 \\
Code translation & 111.33 & 101.74 & 107.54 & 39.80 & 4.41 \\
Bug fixing & 211.72 & 363.59 & 262.24 & 72.23 & 2.80 \\
Mutant generation & 114.80 & 17.89 & 91.95 & 48.83 & 2.54 \\
Assert generation & 139.91 & 91.82 & 129.99 & 55.94 & 5.06 \\
Code summarization & 121.34 & 269.99 & 143.31 & 32.81 & 56.34 \\
Code generation & 231.05 & 384.03 & 240.96 & 78.14 & 7.55 \\
\bottomrule
\end{tabular}
}
\end{table}

Table \ref{tab:tokens_saved} presents the mean number of tokens saved per prompt and an aggregate average. Code generation shows the highest token savings overall (Aggregate: 10,733.42), especially for Llama (8837.50), Qwen (9278.82), and DeepSeek (8502.98). Bug fixing also has very high token savings (Aggregate: 8306.71), 
with DeepSeek (8744.93), Llama (7621.01), and Qwen (7698.30). However, o3-mini lags far behind (360.08). Code QA achieves an exceptionally high savings with Llama (19485.09), much higher than any other model on any task, which may indicate unusually long original prompts or strong compressibility. o3-mini consistently shows the lowest token savings across most tasks, suggesting its prompts may already be concise or less responsive to compression. Tasks with lower overall savings include Assert generation (Aggregate: 250.75), where DeepSeek and Llama still achieve large reductions (3886.71 and 5269.65, respectively), but Qwen and o3-mini do not. Mutant generation also shows lower savings (Aggregate: 2945.01), where Qwen performs particularly poorly (92.55).


A similar shift occurs when response time is considered. Techniques such as \textit{ES-KNN} and \textit{SA} are frequently among the fastest, while approaches like \textit{USC} and \textit{SR} consistently fall to the bottom due to the computational overhead associated with their iterative reasoning or the generation of multiple responses. Despite performance gain in several code generation tasks such as \textit{Code QA} and \textit{Code Generation}, they tend to incur additional time costs to obtain the responses. 

Table \ref{tab:time_saved} presents the average number of seconds saved per prompt. Similar to token savings, Code Generation (231.05s) and Bug Fixing (211.72s) achieved the largest average time savings across models, reinforcing that longer and more complex tasks benefit most from prompt compression. The tasks that benefit least from prompting across models are Exception type prediction (66.68s), Clone detection (80.31s), and Code QA (85.73s). This suggests that prompting techniques provide less efficiency gain for tasks that may be more recognition- or classification-based, rather than generative in nature.




\begin{mdframed}[backgroundcolor=white, linewidth=1pt, linecolor=black]
    \textbf{Observation 4}: 
    Prompting techniques like ES-KNN deliver the highest performance and speed but include more tokens in the prompts, while techniques such as USC boost performance but consume more time to obtain a response. 
\end{mdframed}

\begin{table*}
\centering
\normalsize
\caption{Resource costs associated with Best performing techniques (default / token / time)}
\label{tab:best_cost_benefit_prompting_techniques}
\renewcommand{\arraystretch}{1.2}
\resizebox{\linewidth}{!}{
\begin{tabular}{lcccccc}
\toprule
\textbf{Task} & \textbf{Aggregate} & \textbf{Qwen} & \textbf{DeepSeek} & \textbf{Llama} & \textbf{o3-mini} \\
\midrule
\textbf{Understanding Tasks} & ES-KNN / Control / ES-KNN & ES-KNN / Control / Control & ES-KNN / Control / SA & ES-KNN / RP / ES-KNN & RP / Control / Control \\
Defect detection & ToT / Control / ES-KNN & ToT / Control / Control & ToT / Control / SA & ToT / RP / ES-KNN & RP / Control / Control \\
Clone detection & ES-KNN / SP / ES-KNN & ES-KNN / RP / Control & ES-KNN / EP / TroT & ES-KNN / EP / ES-KNN & ES-KNN / Control / Control \\
Exception type & ES-KNN / Control / ES-KNN & ES-KNN / Control / ES-KNN & ES-KNN / RP / TroT & ES-KNN / Control / ES-KNN & RR / Control / Control \\
Code QA & USC / Control / ES-KNN & SG-ICL / Control / ES-KNN & USC / Control / ES-KNN & USC / RP / ES-KNN & RP / Control / Control \\
\midrule
\textbf{Generation Tasks} & ES-KNN / Control / ES-KNN & ES-KNN / Control / ES-KNN & ES-KNN / RP / SA & ES-KNN / Control / ES-KNN & ES-KNN / Control / Control \\
Code translation & ES-KNN / Control / ES-KNN & ES-KNN / Control / ES-KNN & ES-KNN / SP / SA & ES-KNN / Control / ES-KNN & ES-KNN / Control / Control \\
Bug fixing & Control / Control / ES-KNN & SG-ICL / Control / Control & SG-ICL / RP / TroT & SG-ICL / Control / ES-KNN & Control / Control / Control \\
Mutant generation & ES-KNN / Control / ES-KNN & RP / Control / ES-KNN & ES-KNN / RP / SA & ES-KNN / Control / ES-KNN & RP / Control / Control \\
Assert generation & ES-KNN / Control / ES-KNN & ES-KNN / Control / ES-KNN & ES-KNN / Control / ES-KNN & ES-KNN / Control / ES-KNN & ES-KNN / Control / Control \\
Code summarization & Control / Control / Control & SG-ICL / Control / Control & ES-KNN / Control / ES-KNN & Control / Control / Control & SG-ICL / Control / Control \\
Code generation & USC / Control / ES-KNN & USC / Control / ES-KNN & SG-ICL / SA / ES-KNN & USC / Control / ES-KNN & USC / Control / Control \\
\bottomrule
\end{tabular}
}
\end{table*}

\begin{table*}
\centering
\caption{Resource costs associated with Worst performing techniques (default / token / time)}
\label{tab:worst_cost_benefit_prompting_techniques}
\renewcommand{\arraystretch}{1.2}
\resizebox{\linewidth}{!}{
\begin{tabular}{lcccccc}
\toprule
\textbf{Task} & \textbf{Aggregate} & \textbf{Qwen} & \textbf{DeepSeek} & \textbf{Llama} & \textbf{o3-mini} \\
\midrule
\textbf{Understanding Tasks} & SR / ES-KNN / SR & SR / SR / SR & CCoT / USC / ToT & SR / USC / SR & SBP / SG-ICL / SR \\

Defect detection & EP / SR / SR & RR / SBP / SR & EP / SR / SBP & EP / ES-KNN / SR & SBP / SBP / SR \\
Clone detection & SR / SR / SBP & SG-ICL / SR / SR & SG-ICL / USC / SBP & USC / USC / SR & SR / SR / SBP \\
Exception type & RR / SBP / SBP & SR / SBP / SR & ToT / SBP / SBP & SR / SR / SR & SR / SBP / SR \\
Code QA & TroT / SR / SR & TroT / USC / SR & TroT / USC / SR & Control / TroT / USC & Control / SG-ICL / SG-ICL \\
\midrule
\textbf{Generation Tasks} & SG-ICL / SBP / SG-ICL & SG-ICL / TroT / SR & ToT / SBP / SBP & SG-ICL / SG-ICL / SG-ICL & SR / USC / USC \\
Code translation & SG-ICL / SBP / SG-ICL & SG-ICL / SG-ICL / USC & USC / SBP / SBP & TroT / SG-ICL / SG-ICL & USC / USC / USC \\
Bug fixing & RR / SR / SR & ToT / SBP / SR & RR / SR / SR & SR / SR / SR & SA / USC / USC \\
Mutant generation & ToT / SG-ICL / SG-ICL & ToT / USC / USC & SG-ICL / SBP / ToT & USC / SG-ICL / SG-ICL & SR / SR / SR \\
Assert generation & SR / SG-ICL / SG-ICL & TroT / TroT / TroT & SG-ICL / SG-ICL / SG-ICL & USC / TroT / SR & SR / SR / SR \\
Code summarization & RR / ToT / ToT & RR / ToT / ToT & RR / ToT / ToT & SA / ToT / ToT & RR / USC / ToT \\
Code generation & ES-KNN / SR / SR & TroT / SR / SR & TroT / SR / SR & TroT / TroT / SR & ES-KNN / USC / USC \\
\bottomrule
\end{tabular}
}
\end{table*}

\section{Discussion}
\label{sec:Discussion}



Our results show that prompting effectiveness varies by task and LLM, with no single technique consistently outperforming others. Exemplar Selection KNN was strong across tasks, while Role Prompting offered a cost-efficient alternative. Notably, the \emph{o3-mini} model showed different performance patterns. These findings have several implications. First, they highlight the importance of prompt-model alignment, suggesting that prompt engineering should be adaptive, not one-size-fits-all. Practitioners should empirically validate prompting strategies when switching models, even within the same model family. Second, the variation in performance across tasks reinforces the need for task-aware prompting, where the design of the prompt accounts for the structure, complexity, and requirements of the underlying task. Finally, the deviations seen with \emph{o3-mini} raise concerns about prompt robustness across model scales. This suggests that lighter-weight models may require distinct prompting techniques or additional tuning to match the effectiveness seen in larger models. These insights emphasize the value of prompt evaluation frameworks that account for model-task combinations and point to future research opportunities in automated prompt selection methods.

Our contrastive explanation analysis revealed that prompting techniques like Exemplar Selection KNN outperform others by offering clear structural guidance and relevant in-context examples, enhancing performance across most SE tasks. These findings indicate that effective prompt design requires more than wording—it demands well-structured guidance and tailored examples aligned with the task. Prioritizing prompt clarity and minimizing ambiguity can boost model robustness and efficiency, helping practitioners develop more effective prompting strategies to maximize LLM performance in diverse software engineering tasks. Moreover, these insights suggest that adaptive prompt engineering, which dynamically incorporates task-specific structures and examples, could further improve results. This also highlights the potential for automated prompt optimization tools to assist developers in crafting prompts that align with task requirements, ultimately accelerating development workflows and reducing trial-and-error in prompt design.


Our analysis of token savings and time savings reveals that prompt optimization offers significant efficiency gains across SE tasks, with clear variation based on task type and LLM architecture. Tasks such as Code Generation and Bug Fixing exhibited the highest average token savings, exceeding 8,000 tokens per prompt. This suggests that these token-intensive tasks benefit substantially from prompt compression strategies, with potential reductions in inference cost and latency. These are critical considerations in real-world applications that utilize commercial LLMs with token-based billing.

We also observed considerable variation in token and time savings across LLMs. Larger models like Llama and DeepSeek consistently achieved higher savings, indicating they are more amenable to prompt compression, possibly due to differences in tokenization schemes, context window capacity, or internal architectural design. In contrast, o3-mini showed relatively limited token savings across most tasks, suggesting a constrained capacity for compression. These findings reinforce that prompt optimization strategies should be model-aware, as techniques effective on one LLM may not generalize to another.

Task-specific trends also emerged. For example, Assert Generation and Exception Type classification yielded lower aggregate token savings, implying that prompts for these tasks may already be concise, or that further compression risks omitting critical context. Conversely, Code QA with Llama showed exceptionally high savings, warranting further investigation into whether such tasks include redundancies or patterns that facilitate aggressive but safe compression. Some tasks, like Exception Type and Clone Detection, consistently had lower time savings across models, suggesting their input prompts may already be compact or that compression has limited impact on runtime.

From a practical standpoint, these findings emphasize that prompt design should carefully balance efficiency and performance. While reducing token usage is desirable, it must not come at the expense of output quality, particularly for complex tasks like Bug Fixing and Code Translation, which may require more detailed and nuanced input. The observed differences across tasks and models highlight an opportunity for automated, context-aware prompt rewriting tools that tailor prompt structure dynamically to maximize utility and efficiency.

\section{Threats to Validity}
\label{sec:Threats}
In this section, we outline the potential threats to the validity of our study.

\textbf{Construct Validity:}
To mitigate bias in manual evaluation for prompting technique selection (Section~\ref{sec:Defining_Prompting_Techniques} and prompt variation template selection (Section~\ref{sec:prompt_validation}), the researchers independently conducted the manual analysis. Following open-coding practices~\cite{Glaser2016OpenCD}, discrepancies were resolved through negotiated agreement. Moreover, to ensure the quality of model-generated responses through contrastive explanations, we adhered to established prompting practices~\cite{jacovi2021contrastive}.

 
\textbf{Internal Validity:}
To ensure that task performance reflects solely the effectiveness of the prompting technique rather than prompt wording or model temperature, we used a fixed default temperature value (1 for o3-mini and Deepseek, 0.7 for Llama and Qwen) for all LLM and we employed the prompt validation process as outlined in Section~\ref{sec:prompt_validation}. In addition, we adopted the dataset used by a prior study~\cite{The_Prompt_Report}. We used the same sampled dataset for each SE task to compare the performance of different prompting techniques across four LLM models. 


\textbf{External Validity:}
This study uses four LLMs to investigate 10 software engineering (SE) tasks and 14 prompting techniques. While our findings lay a foundation for the systematic evaluation of prompting techniques across diverse SE tasks, they may not be fully generalizable beyond the specific tasks, datasets, prompting techniques, and LLMs used in this work. To minimize the threat pertaining to LLMs, we selected models with distinct training objectives (general-purpose vs. code-specialized) and model availability (open-source vs. proprietary models) and strong performance on the widely-used EvalPlus leaderboard~\cite{10.5555/3666122.3667065}.

\section{Conclusion and Future Work}
\label{sec:Conclusion}

In this study, we conducted systematic evaluation of 14 prompt engineering techniques across 10 SE tasks using four LLMs. Our results offer concrete insights into \emph{which} techniques yield highest performance gains and \emph{where} resource overheads may present practical limitations in real-world deployments. The findings of our work provide empirical evidence to guide practitioners and researchers for selecting optimal prompting techniques that are best aligned with task-specific objectives and real-world operational constraints such as execution time and token usage. 


Future work should extend beyond the linguistic characteristics of natural language prompts to explore additional dimensions, such as properties of prompt embeddings. Further research is needed to investigate strategies to optimize the prompt structure in accordance with ask-specific requirements and dataset characteristics. All research artifacts are available on our companion website.~\cite{replication_package}

\bibliographystyle{IEEEtran}
\bibliography{sample-base}


\end{document}